\documentstyle[twoside]{article}
\newcount\Mac  \Mac=1  
\newcommand{\ifMac}[2]{\ifnum\Mac=1 #1 \else #2 \fi}
\newcommand{\riga}[1]{\noalign{\hbox{\parbox{\textwidth}{#1}}}\nonumber}

\newcommand{\GeV}{\,{\rm GeV}}
\newcommand{\TeV}{\,{\rm TeV}}

\newcommand{\PRL}{Phys. Rev. Lett.}
\newcommand{\PL}{Phys. Lett.}
\newcommand{\PR}{Phys. Rev.}

\newcommand{\LH}{\Lambda_{\hbox{\footnotesize H\tiny ERA}}}

\newcommand{\mb}[1]{\mbox{\normalsize\boldmath $#1$}}
  \def\SU{{\rm SU}}
 
\def\circa#1{\,\raise.3ex\hbox{$#1$\kern-.75em\lower1ex\hbox{$\sim$}}\,}
\newcommand{\eq}[1]{~{\rm (\ref{eq:#1})}}
\newcommand{\sys}[1]{~{\rm (\ref{sys:#1})}}

\def\Red{}
\def\Black{}
\def\Blue{}

\newcommand{\lascia}[1]{}
\makeatletter
\def\art{\@ifnextchar[{\eart}{\oart}}
\def\eart[#1]#2#3#4#5#6{{\rm #2}, {\em #3 \bf #4} {\rm (#6) #5}}
\def\hepart[#1]#2{{\rm #2, \em#1}}
\newcommand{\oart}[5]{{\rm #1}, {\em #2 \bf #3} {\rm (#5) #4}}
\newcommand{\y}{{\rm and} }

\newcounter{alphaequation}[equation]
\def\thealphaequation{\theequation\hbox to
0.6em{\hfil\Alph{alphaequation}\hfil}}
\def\eqnsystem#1{
\def\@eqnnum{{\rm (\thealphaequation)}}
\def\@@eqncr{\let\@tempa\relax \ifcase\@eqcnt \def\@tempa{& & &} \or
  \def\@tempa{& &}\or \def\@tempa{&}\fi\@tempa
  \if@eqnsw\@eqnnum\refstepcounter{alphaequation}\fi
\global\@eqnswtrue\global\@eqcnt=0\cr}
\refstepcounter{equation} \let\@currentlabel\theequation \def\@tempb{#1}
\ifx\@tempb\empty\else\label{#1}\fi
\refstepcounter{alphaequation}
\let\@currentlabel\thealphaequation
\global\@eqnswtrue\global\@eqcnt=0 \tabskip\@centering\let\\=\@eqncr
$$\halign to \displaywidth\bgroup \@eqnsel\hskip\@centering
$\displaystyle\tabskip\z@{##}$&\global\@eqcnt\@ne
\hskip2\arraycolsep\hfil${##}$\hfil& \global\@eqcnt\tw@\hskip2\arraycolsep
$\displaystyle\tabskip\z@{##}$\hfil
\tabskip\@centering&\llap{##}\tabskip\z@\cr}

\def\endeqnsystem{\@@eqncr\egroup$$\global\@ignoretrue} \makeatother

\begin{document}
\twocolumn[
\begin{quote}{\em 10 June 1997\hfill \bf hep-ph/9706298}\\
{\bf SNS/PH/1997-007 \hfill IFUP--TH/23-96}\end{quote}
 \vspace{1cm}
\centerline{\huge\bf\Red Atomic parity violation}
\centerline{\huge\bf and the H{\LARGE\bf ERA} anomaly}

\bigskip\bigskip\Black
\centerline{\large\bf Leonardo Giusti} \vspace{0.2cm}
\centerline{\em Scuola Normale Superiore, Piazza dei Cavalieri 7}
\centerline{\em {\rm and} INFN, sezione di Pisa,  I-56126 Pisa, Italia}
\vspace{5mm}\centerline{\large and}\vspace{5mm}
\centerline{\large\bf Alessandro Strumia}\vspace{0.2cm}
\centerline{\em Dipartimento di Fisica, Universit\`a di Pisa {\rm and}}
\centerline{\em INFN, sezione di Pisa,  I-56126 Pisa, Italia}\vspace{1cm}
\Blue

\centerline{\large\bf Abstract}
\begin{quote}\large\indent
We show that the two scenarios able to explain the {\sc Hera\/} anomaly ---
a new leptoquark coupling or a new contact interaction ---
predict new contributions to atomic parity violation.
These corrections are sufficiently large and different that
a feasible reduction in the dominant atomic theory uncertainty
could give some hint in favour of one of the two scenarios.
\end{quote}\Black
\vspace{1cm}]


\paragraph{1}
The excess of events at large $Q^2$ observed at {\sc Hera} \cite{Hera}
could be due to new physics.
In this case two different mechanisms can
account for the {\sc Hera} anomaly without contradicting present
experimental bounds:
\begin{itemize}
\item an effective four-fermion interaction~\cite{Contact,RC},
related to a new physics scale or
produced by an exchange of a particle in the $t$-channel;
\item
the resonant $s$-channel exchange
of a particle with `leptoquark' couplings~\cite{RC,R,L}, that would give
a characteristic peaked distribution for the invariant mass
of the final states.
\end{itemize}
The aim of this work is to show that 
the new physics would give a non negligible predictable
extra contribution to atomic parity violation (APV)
in both cases.
This happens as follows.

\smallskip

In the first scenario,
the new contact interaction does not
contribute resonantly to the {\sc Hera} process, and
affects significantly other physical observables.
In particular, the correction to APV is the most interesting one:
if a {\em single\/} new operator should explain the {\sc Hera} anomaly,
the correction to APV would be $(10\div 20)$ times
larger than the present error on its determination.
It is thus necessary to assume the existence of different contact operators
which individually give a too large contribution to APV,
but related in such a way that the total contribution cancels.
Such cancellation, at tree-level, can be justified
invoking appropriate global symmetries~\cite{SU12}.
As discussed in section~2, however,
radiative corrections from the standard
gauge interactions upset the cancellation
giving a predictable correction to APV.

In the second scenario the leptoquark exchange
gives a resonantly enhanced contribution to the {\sc Hera} process
and a contribution to APV~\cite{RC,DeAndrea} that is
not completely negligible,
as discussed in section~3.

\smallskip

Before discussing these points in more detail,
let us summarize the present determination of APV.
The `weak charge' $Q_{\rm W}$
that parametrizes
the parity-violating effective Ha\-miltonian~\cite{Sap}
$${\cal H}_{\rm APV}=\frac{G_{\rm F}}{2\sqrt{2}}
Q_{\rm W} \rho_{\rm nucleus}(\mb{r}) \gamma_5$$
that dominates APV in cesium,
is predicted by the SM (constrained by the LEP data) to be~\cite{PdG,SMAPV}
$$Q_{\rm W}|_{\rm SM} = -73.17\pm0.13,$$
and has recently been measured to be~\cite{QWexp}
$$Q_{\rm W}|_{\rm exp} = -72.11\pm(0.27)_{\rm exp}\pm(0.89)_{\rm th}.$$
The dominant theoretical error is due to
uncertainty in the atomic wave function:
even in the most favourable case of cesium this error is about $1\%$.
A lengthy computation based on an expansion in
the number of excited electrons can however
reduce significantly this theoretical error, maybe down to the
few {\em per mille\/} level~\cite{Sap}.
Furthermore, the accuracy of this computation can be tested
comparing its results with some accurately measured properties of cesium,
like the hyperfine constants and the energy levels~\cite{Sap}.

\smallskip

The main result of this work, summarized in the conclusion,
is that a reduction of the uncertainty on $Q_{\rm W}$
could reasonably give interesting indications
about the nature of the (eventual) new physics
suggested by the {\sc Hera} data.

\paragraph{2}
In this section we consider the case where the {\sc Hera} anomaly is produced
by contact interactions, present in combinations that, at tree level,
do not contribute to APV.
Even in this case, a contribution to APV arises because
weak radiative corrections do not respect the
cancellation between different contributions
(or the symmetry at the basis of the cancellation).
Since the contribution of each individual term is quite large,
$\Delta Q_{\rm W}\sim (10\div 20)$~\cite{Contact,RC},
the radiatively generated effect\footnote{The first
experiments on APV gave
a value smaller than the prediction of the SM,
stimulating the computation of these kind of corrections~\cite{Marciano}
in models with no APV at tree level, like in
$\SU(2)_L\otimes\SU(2)_R\otimes{\rm U}(1)$.},
that only depends on the operator structure
and on the standard gauge interactions,
is expected to be numerically interesting.
For example, in the SM~\cite{SMAPV}, there is a $\approx 5\%$ correction to the
tree level SM contribution
$Q_{\rm W}=-N+Z(1-4\sin^2\theta_{\rm W})$.

\medskip

There are two possible combinations ``A'' and ``B'' of gauge-invariant
contact operators~\cite{Contact,RC}
able to explain the {\sc Hera} anomaly without contradicting
other bounds from LEP and {\sc TeVATRON}.
They are
described by the following additional terms in the Lagrangian:
\begin{eqnsystem}{sys:AB}\label{eq:A}
{\cal L}_{\rm eff}^{\hbox{\footnotesize ``A''}}
&=& \frac{4\pi}{\LH^2}(\bar{e}_R\gamma_\mu e_R)\bigg\{
(\bar{Q}\gamma^\mu Q)+\\
&&-(\bar{u}_R \gamma^\mu u_R)-(\bar{d}_R \gamma^\mu d_R)\bigg\}\nonumber \\
\riga{in case ``A'', and}\\[-2mm]
\quad{\cal L}_{\rm eff}^{\hbox{\footnotesize ``B''}} &=&
\frac{4\pi}{\LH^2}\bigg\{(\bar{e}_R\gamma_\mu e_R)
(\bar{Q}\gamma^\mu Q)+\\
&&\nonumber+
(\bar{L} \gamma_\mu L)[(\bar{u}_R \gamma^\mu u_R)+(\bar{d}_R \gamma^\mu d_R)]\bigg\}
\end{eqnsystem}
in case ``B''.
Here $Q,u_R,d_R,L,e_L$ are the standard notations for the SM matter fermions.
The relative sign between the various operators is fixed to
cancel the contributions to APV;
the overall sign is fixed by the necessity of having a positive
interference with the SM contribution to the {\sc Hera} process~\cite{Contact,RC}.

The structure of the operators, seemingly very artificial,
can receive some partial theoretical justification.
For example, the quark current in\eq{A} can be forced to be axial
(so that the contribution to $Q_{\rm W}$ vanishes)
imposing an SU(12) symmetry
acting on the quark fields~\cite{SU12}.
Such a symmetry could naturally arise in composite models.
However, considerations of this kind do not explain the
suppression of other, possible but unwanted, contact interactions,
for example involving leptons only.

\medskip

\begin{table}\Blue
$$\begin{array}{|c|cc|}\hline
\hbox{correc-}&\multicolumn{2}{|c|}{\hbox{contact operators}}\\
\hbox{tion to}&\hbox{``A''} & \hbox{``B''}\\ \hline\hline
C_{\rm e.m.} & 12Z &0\\
C_Y & 9Z+3N &\frac{1}{3}(2Z+11N)\\ \hline
\Red\Delta Q_{\rm W}\Blue & \Red+0.72 &+0.15\Blue\\ \hline
\end{array}$$\Black
\caption[24]{\em Correction to parity violation in atoms in
the two scenarios {\rm ``A''} and {\rm ``B''}
of contact interactions.
The coefficients $C_i$ are defined by eq.\eq{dQW}, $Z$ is
the atomic number and $N$ is the number of neutrons.}
\end{table}

The computation of the renormalization corrections to
the various contact operators is lengthy but
straightforward.
The leading logarithmic
correction to APV produced by SM gauge interactions in the two interesting cases
``A'' and ``B'' are:
\begin{eqnarray}\nonumber
\Delta Q_{\rm W}&=& \frac{4\pi}{\sqrt{2}G_{\rm F}\LH^2}\bigg\{
C_{\rm e.m.} \frac{\alpha_{\rm e.m.}}{4\pi}\ln\frac{M_Z^2}{\Lambda_{\rm QCD}^2}+\\
&&+
C_Y \frac{\alpha_Y}{4\pi}\label{eq:dQW}
\ln\frac{\LH^2}{M_Z^2}\bigg\}
\end{eqnarray}
where we have used standard notations for the various quantities,
in particular
$\Lambda_{\rm QCD}\approx 300\GeV$ is
the low-energy soft mass term of light quarks, and
the numerical coefficients $C_i$ are given in table~1.
Their computation is simplified noticing that many diagrams
give no contribution.
Strong interactions cannot upset the cancellation.
In our approximation, $\SU(2)_L$ gauge interactions
do not contribute
due to gauge invariance and to the absence of operators
involving both lepton and quark doublets.
`Penguin' diagrams can give a contribution only when mediated
by the hypercharge gauge boson,
and vanish in case ``A'' because the quark current is axial.

Assuming that
$\LH=3\TeV$~\cite{Contact,RC} so that the new interactions
can account for {\sc Hera} anomaly, the correction
to the weak charge in cesium
($Z=55$ and $N=58$) is\footnote{Another possible
operator ``C''~\cite{Contact,RC}
$(4\pi/\LH^2)\*(\bar{L}\gamma_\mu L)\*
(\bar{u}_R\gamma^\mu u_R-\bar{d}_R \gamma^\mu d_R)$
is accompanied by a much larger effect $\Delta Q_{\rm W}\approx - 3$
(given by a tree level term, proportional to $N-Z$,
plus a radiatively generated contribution).}
\begin{equation}\label{eq:dQAB}
\Delta Q_{\rm W}|_{\hbox{\footnotesize ``A''}} = +0.72,\qquad
\Delta Q_{\rm W}|_{\hbox{\footnotesize ``B''}} = +0.15.
\end{equation}
In both cases ``A'' and ``B'' the contribution
to the smaller spin-dependent APV effects
remains below the uncertainty due to QCD effects.

The result\eq{dQAB} is not much different in {\em supersymmetric models}
for the contact interactions.
In these cases it is natural to assume that contact interactions arise as
supersymmetric $D$-term operators, like
the supergauge-invariant extension of
$\int d^4\!\theta \,\hat{e}^\dagger\hat{e} \,\hat{q}^\dagger \hat{q}$,
where $\theta$ is the superspace parameter and $\hat{e}$ and $\hat{q}$ are
lepton and quark superfields.
The computation of the full supersymmetric `supergauge' corrections,
i.e.\ the inclusion of gaugino-sfermion loops, can be conveniently
done via superfield techniques
(of course, the fact that now we employ the supersymmetric Feynman
gauge does not affect the gauge-invariant correction to atomic parity violation).
Assuming that the various squarks of first generation have a common mass
$m_{\tilde{q}}$,
the contribution to $Q_{\rm W}$ in eq.\eq{dQW} from operators ``A''
at energies between $\LH$ and
$\Lambda_{\rm SUSY}=\max(m_{\tilde{q}},m_{\tilde{e}}, M_{\tilde{B}})$
has a modified coefficient $C_Y^{\rm SUSY}=6Z+2N$, giving
a slightly reduced contribution to $Q_{\rm W}$.
We have indicated with $m_{\tilde{e}}$ a generic selectron mass
and with $M_{\tilde{B}}$ the bino mass.
In case ``B'' there are contributions from supersymmetric penguins,
so that two distinct supersymmetric masses,
one for quarks and one for leptons, should be introduced.
Assuming, for simplicity, that the two masses are equal,
we find a value $C_Y^{\rm SUSY}=3N+Z$
almost numerically identical to the non supersymmetric case.

\paragraph{3}
In the leptoquark scenario the correction to atomic parity violation
is somewhat smaller than its present accuracy.
Since this point has already been discussed in~\cite{RC,DeAndrea},
we will concentrate on a specific and motivated realization of this scenario,
in which the APV effect turns out to be interesting.
Supersymmetric theories furnish a theoretically more motivated
realization of the leptoquark scenario that automatically
offers `invisible' channels for the leptoquark ($LQ$) decay,
$B\equiv\hbox{B.R.}(LQ\to e q)<1$
(as suggested by {\sc TeVATRON} bounds~\cite{RC}).
More precisely, in the context of supersymmetry,
introducing
`$R$-parity violating' supersymmetric interactions~\cite{MSSMR},
a squark $\tilde{Q}_3$ can have the `leptoquark' interaction
$$
\lambda'_{131} \bar{L}_1 \tilde{Q}_3 d_{R1}+\hbox{h.c.}
$$
able of producing the {\sc Hera} anomaly if a stop state is
sufficiently light
(here 131 are generation indices\footnote{A leptoquark of second generation,
$\tilde{Q}_2$,
or an interaction with a `sea' quark, $s_R$,
are less interesting, but not excluded, alternative possibilities~\cite{R}.}).
In supersymmetry a (mainly right-handed) stop state can naturally be
lighter than the gluino and the other squarks.
In this scenario the contribution to the APV parameter $Q_{\rm W}$ is
$$\Delta Q_{\rm W} = -\frac{|\lambda'_{131}|^2(2N+Z)}{2\sqrt{2}G_F}\left[
\frac{\sin^2\theta_{\tilde{t}}}{m_{\tilde{t}}^2}+
\frac{\cos^2\theta_{\tilde{t}}}{m_{\tilde{T}}^2}\right].$$
Here $\theta_{\tilde{t}}$ is the $L$eft/$R$ight stop mixing angle
($\theta_{\tilde{t}}=0$ for a purely right-handed lighter stop),
$m_{\tilde{t}}$ is the mass of the lighter stop $\tilde{t}$
that gives rise to the {\sc Hera} anomaly
and $m_{\tilde{T}}$ is the mass of the heavier stop $\tilde{T}$.
Inserting the values suggested by the {\sc Hera} data~\cite{Hera},
$$m_{\tilde{t}}\approx 200\GeV,\qquad 
|\lambda'_{131}\sin\theta_{\tilde{t}}|\approx
\frac{0.04}{\sqrt{B}}\quad\cite{RC,R},$$
the correction to the weak charge in Cesium is
\begin{equation}\label{eq:QWstop}
\Delta Q_{\rm W} = -\frac{0.26}{B}\left(1+\frac{1}{\tan^2\theta_{\tilde{t}}}
\frac{m_{\tilde{t}}^2}{m_{\tilde{T}}^2}\right).
\end{equation}
Naturalness considerations suggest that
$|\theta_{\tilde{t}}|\circa{<}0.3$
(a light left-handed stop would also give a
too large electroweak correction to $M_Z/M_W$) and
$m_{\tilde{T}}\circa{<}500\GeV$~\cite{LRstop}, so that,
including the contribution from the heavier stop,
one finds a result $2\div3$ times larger than the `naive' one.

\paragraph{4}
In conclusion, the most appealing scenarios able of explaining the {\sc Hera}
anomaly predict the following non-SM contribution $\Delta Q_{\rm W}$
to the weak charge $Q_{\rm W}$ that gives the dominant
parity-violating effect in Cesium:
\begin{eqnarray*}\Blue
\parbox{3.5cm}{\centering contact operators\\ ``A'' and ``B'' in\sys{AB}}&&
\hspace{-0.2cm}\begin{array}{l}
\Delta Q_{\rm W}|_{\hbox{\footnotesize ``A''}}\approx +(0.6\div 0.7)\\
\Delta Q_{\rm W}|_{\hbox{\footnotesize ``B''}}\approx +0.15
\end{array} \\[3mm]
\parbox{3.5cm}{\centering light stop with\\ leptoquark couplings}
&&\Delta Q_{\rm W}|_{\rm stop}\circa{<}- 0.6\Black
\end{eqnarray*}
These results should be compared with
$$\Blue Q_{\rm W}|_{\rm exp} - Q_{\rm W}|_{\rm SM}=
+1.06\pm(0.30)_{\rm exp}\pm(0.89)_{\rm th}\Black$$
As mentioned above, a lengthy but feasible and checkable atomic theory computation can
reduce the dominant theoretical uncertainty
below the experimental one~\cite{Sap}.
In this case, depending on the new central value of $Q_{\rm W}$
(for example if would remain unchanged),
one could obtain some hint in favour
of one of the possible scenarios.

\paragraph{Acknowledgments}
We thank R.\ Barbieri for helpful discussions and
for having instructed us about gauge invariance.

\end{document}